\documentclass[prd,reprint,amsmath,amssymb,floatfix,aps,longbibliography,superscriptaddress,nofootinbib]{revtex4-2}

\usepackage{amsmath,graphicx,amssymb,xcolor,booktabs,braket,multirow,float,cancel,enumerate,enumitem}
\usepackage[colorlinks=true, allcolors=blue]{hyperref}
\usepackage[normalem]{ulem}
\usepackage{soul}

\setstcolor{red}

\definecolor{Mahogany}{rgb}{0.75, 0.25, 0.0} 

\definecolor{Plum}{rgb}{0.56, 0.27, 0.52}

\begin{document}

\title{Gemini dark matter}

\author{Andrew Cheek}
\email{acheek@sjtu.edu.cn}
\affiliation{Tsung-Dao Lee Institute, Shanghai Jiao Tong University, Shanghai, 201210, China}
\affiliation{Key Laboratory for Particle Astrophysics and Cosmology (MOE) \& Shanghai Key Laboratory for Particle Physics and Cosmology, Shanghai Jiao Tong University, Shanghai 200240, China}

\author{Yu-Cheng Qiu}
\email{ethanqiu@sjtu.edu.cn}
\affiliation{Tsung-Dao Lee Institute, Shanghai Jiao Tong University, Shanghai, 201210, China}

\author{Liang Tan}
\email{tanliang@sjtu.edu.cn}
\affiliation{Tsung-Dao Lee Institute, Shanghai Jiao Tong University, Shanghai, 201210, China}
\affiliation{Key Laboratory for Particle Astrophysics and Cosmology (MOE) \& Shanghai Key Laboratory for Particle Physics and Cosmology, Shanghai Jiao Tong University, Shanghai 200240, China}

\date{\today}

\begin{abstract}
The $S_8/\sigma_8$ tension in the large-scale structure can be explained by decaying dark matter with an almost degenerate spectrum and small enough decay width.
Here we propose the Gemini dark matter model, which contains a heavy mother particle $\chi_3$ and two twins $\chi_{1/2}$, which are almost degenerate in mass and are produced at the same time. 
The dark sector is charged under the same Froggatt-Nielsen symmetry that can explain the hierarchy of the Standard Model Yukawa couplings.
The slightly heavier $\chi_2$ decays into $\chi_1$ and the axionic component of the flavon, which washes out the small-scale structure and resolves $S_8/\sigma_8$ tension. We present the production mechanism of Gemini dark matter and viable parameter regions. We find that despite the preferred dark matter mass being $\mathcal{O}(1)$--$\mathcal{O}(100)$ keV, they constitute cold dark matter.
The Gemini dark matter model predicts an abundance of dark radiation that will be probed in future measurements of the cosmic microwave background.

\end{abstract}

\maketitle

\section{Introduction}\label{sec:intro}

The standard model of modern cosmology, the $\Lambda$CDM model, currently provides a consistent picture of the observable Universe to a satisfying level of precision~\cite{Planck:2018vyg}. In recent years, however, some observations have put tension on the $\Lambda$CDM~\cite{Abdalla:2022yfr},
such as the Hubble tension and the $S_8/\sigma_8$ tension~\cite{DiValentino:2020vvd}. If either of these tensions persist, cosmologists will have to 
go beyond $\Lambda$CDM. This paper focuses on a new physics solution to the $S_8/\sigma_8$ tension. 

Decaying dark matter (DDM)~\cite{Turner:1984ff} 
has been proposed to resolve the $S_8/\sigma_8$ tension~\cite{Aoyama:2014tga,Enqvist:2015ara,FrancoAbellan:2020xnr,FrancoAbellan:2021sxk,Simon:2022ftd,Fuss:2022zyt,Bucko:2023eix,Fuss:2024dam}. 
The basic solution states that a cold dark matter (CDM) decays into a slightly lighter dark particle. The mass difference gives the product particle some velocity so it acts as warm dark matter (WDM), washing out structure a little.
The decay lifetime should be $\tau \sim \mathcal{O}(10)$--$\mathcal{O}(100)~{\rm Gyr}$ and the mass ratio between the decay product and the CDM, quantified as $\epsilon\equiv (m_{\rm CDM}^2 - m_{\rm WDM}^2)/2m_{\rm CDM}^2$, must be $\sim 0.01$--$0.1$~\cite{Fuss:2024dam}.
This almost degenerate dark sector and highly suppressed decay must also evade indirect-detection bounds. This could be achieved by assigning new quantum numbers and discrete symmetries.

The Standard Model (SM) of particle physics also contains several mysteries,
and one of them is
the mass hierarchy between the three generations of quarks and leptons.
One solution is the well-developed Froggatt-Nielsen (FN) mechanism~\cite{Froggatt:1978nt,Leurer:1993gy,Leurer:1992wg}, which introduces a chiral $U(1)_{\rm FN}$ (global or gauge) symmetry to the SM particles. 
This new symmetry forbids Yukawa terms at the renormalizable level. The dimension at which the Yukawa term appears at an effective nonrenormalizable level depends on the FN quantum numbers of the given particle.
To retrieve the SM description, the $U(1)_{\rm FN}$ symmetry is spontaneously broken where the Yukawa couplings are now proportional to 
a common parameter, $\lambda  <1$ to some positive generation-dependent power,
and hierarchy emerges.
This new symmetry is associated with a new scalar field, named the flavon $\Phi$, and it mediates flavor-changing 
current,
whose coupling strengths with fermions are proportional to the respective fermion masses. In other words, flavon couplings are generation dependent. 

The FN framework can be easily extended to the dark sector~\cite{Calibbi:2015sfa}. In the minimal setup, the dark sector interacts with the SM particles by mediating the flavon $\Phi$. 
Cosmologically, DM can be produced through either thermal freeze-out or nonthermal freeze-in~\cite{Calibbi:2015sfa,Cheek:2022yof,Babu:2023zni,Mandal:2023jnv}. Since the coupling between the flavon and the DM particle depends on DM mass, the direct and indirect detection may have interesting signatures. 

Reference~\cite{Cheek:2022yof} considered DDM under the FN framework and 
suggested a potential model to resolve the $S_8$, albeit very roughly. In this paper, we systematically study the mass spectrum required to solve the $S_8$ tension. We have identified a more consistent solution that results in different cosmology and phenomenological signals.
In the dark sector, three dark fermions are charged under the FN symmetry. Two of them, $\chi_{1/2}$, are almost degenerate, and the other one, $\chi_3$, is much heavier, $m_3\gg m_{1/2}$. The decay channel $\chi_2 \to \chi_1 + a$, where $a$ is the axionic component of the complex FN scalar $\Phi$, is responsible for resolving the $S_8/\sigma_8$ tension. The $\chi_{1/2}$ are produced from $\chi_3$ decay, $\chi_3 \to \chi_{1/2} + a$. We name this model ``Gemini dark matter" (Gemini DM hereafter), where particles $\chi_{1/2}$ are the twins and $\chi_3$ is the mother particle. 
The mother particle is produced through 
freeze-in from the SM thermal bath. The preferred twin masses in this model are roughly $\mathcal{O}(1)$--$\mathcal{O}(100)$~keV, and their average velocity is~$\sim 10^{-6}$, which is cold despite its small mass.

The Gemini DM model predicts the existence of the dark radiation relic. During the parturition $\chi_3\to \chi_{1/2}+a$, the light axionic flavon $a$ is also produced as dark radiation, which shall contribute to the deviation from the effective number of neutrino species $\Delta N_{\rm eff}$.
It may be probed in future observations by CMB-S4~\cite{CMB-S4:2016ple} and CMB-HD~\cite{CMB-HD:2022bsz}.

This paper is organized as follows. 
In Sec.~\ref{sec:model} we construct the Gemini DM model. The cosmological production of Gemini DM is explained in Sec.~\ref{sec:production}. 
Summary and discussions are in Sec.~\ref{sec:summary}. 
The Appendix reviews the FN mechanism in the SM sector.

\section{Gemini dark matter}\label{sec:model}

In this section, we construct the Gemini DM model by extending the FN symmetry to the dark sector, which contains Weyl fermions $\chi_i$ carrying FN charge $n_i$, which is also the SM gauge group singlet. There are three generations of $\chi$'s. We focus on constructing a particle spectrum that resolves the $S_8/\sigma_8$ tension, i.e., we require two particles with almost degenerate masses. 

With our setup, below some UV cutoff scale $\Lambda$, the leading-order nonrenormalizable operators that respect the global $U(1)_{\rm FN}$ appearing in the dark sector Lagrangian density are
\begin{equation}
\mathcal{L} \supset i\bar{\chi}_j\bar{\sigma}^\mu \partial_\mu \chi_j -  \frac{\beta_{ij}}{2} \frac{\Phi^{n_i + n_j}}{\Lambda^{n_i+ n_j -1}}\chi_i \chi_j + {\rm H.c.}\;,
\label{eq:interactions}
\end{equation}
where $\beta_{ij}\sim \mathcal{O}(1)$ are the coupling constants and $\Phi$ is the flavon field with FN charge $-1$. Here a bar indicates the conjugate and $\bar{\sigma}^\mu = (1,-\vec{\sigma})$, where $\vec{\sigma}$ is an array of Pauli matrices. 
The summation of dummy indices is employed here.
The complex FN scalar $\Phi$ can acquire a nontrivial vacuum expectation value, $\langle \Phi \rangle = f_a$, under its potential, leading to spontaneous symmetry breaking of $U(1)_{\rm FN}$. Since we are considering the global $U(1)_{\rm FN}$, and quantum gravity cannot accommodate such symmetries, it must be broken at least by the Planck scale~\cite{Banks:2010zn, Reece:2023czb}. Therefore, the Goldstone boson $a$ (we call it the axion for familiarity) acquires a mass through an explicit breaking term.
This axion generally could couple to the QCD gluon field via an anomaly term, depending on SM FN charges. One could identify this axion as precisely the QCD axion~\cite{Calibbi:2016hwq,Ema:2016ops}. However, in this paper, we consider it a general axion-like particle that does not solve the strong $CP$ problem. Therefore, we consider its mass to be a free parameter.\footnote{If we adopt the discrete FN symmetry $Z_{33}$ as mentioned in Ref.~\cite{Qiu:2023igq}, which is anomaly-free, there is no coupling between the axion and the gluon. So the axion mass is a free parameter.}
After spontaneous symmetry breaking, the field can be written as $\Phi = e^{ia/f_a}(f_a+\phi)/\sqrt{2}$. A Taylor expansion gives 
\begin{equation}
    \left(\frac{\Phi}{\Lambda}\right)^n \to \lambda^n e^{i n a/ f_a} \left(1 + \frac{n \phi}{f_a} + \cdots\right)\;,
    \label{eq:expansion}
\end{equation}
where the FN parameter is defined as $\lambda \equiv f_a/\sqrt{2}\Lambda$. 
The leading interaction terms in Eq.\eqref{eq:interactions} are the Yukawa-type interactions that determine the $\chi$ masses. After performing the phase rotation for each fermion, $\chi_j \to e^{-i n_j a/f_a} \chi_j$, one rotates the axion away. 
This phase rotation generates couplings between $\partial a$ and the $\chi$'s through the $\Phi$ field's kinetic term. Using the equation of motion for the $\chi$'s, one obtains the interaction between $a$ and fermions. Due to the mismatch between mass eigenstates and coupling states, flavor-changing currents that couple to the flavon are still present.
After diagonalizing the mass matrix, the interactions in Eq.~\eqref{eq:interactions} are expressed as
\begin{equation}
-\mathcal{L} \supset  \frac{1}{2} m_k \chi_k \chi_k + g_{ij}^\phi \phi \chi_i \chi_j + g_{ij}^a a\chi_i \chi_j + {\rm H.c.}\;,
\label{eq:Lchi}
\end{equation}
where $m_k$ is the $k$th diagonal component of $D$ and
\begin{align*}
D &= {\rm diag}(m_1,m_2,m_3) = U^\top M U \;,\\
g_{ij}^\phi & =  \frac{1}{2}\left[ {\rm sym}\left( U^\top \frac{1}{f_a} \frac{\partial(\lambda  M) }{\partial \lambda} U \right) \right]_{ij}\;, \\ 
g_{ij}^a & = \left[{\rm sym}\left(\frac{N D}{f_a} \right)\right]_{ij} \;, \\
M_{ij} & =\frac{1}{\sqrt{2}} f_a \beta_{ij}\lambda^{n_i + n_j-1} \;, \\
N_{ij} & = (U^\dagger)_{ik} n_k U_{kj}\;.
\end{align*}
The $U$ is a unitary matrix, $UU^\dagger =1$.
Here $[{\rm sym}(A)]_{ij}\equiv A_{ij}+A_{ji}- A_{ii}\delta_{ij}$, for any matrix $A$. 
The presence of this symmetric sum for the off-diagonal couplings in $g_{ij}^{\phi,a}$ is due to the spinor identity $\chi_i\chi_j = \chi_j\chi_i$. 
The nonvanishing flavor-changing couplings between flavons and $\chi$ fields provide decay channels.

The Gemini DM model requires at least three generations of $\chi$. The reason is the following.
In the minimal setup, we require the decay channel $\chi_2\to \chi_1 + a$ to resolve the $S_8$ tension. Meanwhile, the mass split between $\chi_1$ and $\chi_2$ is small, which is described by
\begin{equation}
    \epsilon = \frac{1}{2} \left(1-\frac{m_1^2}{m_2^2}\right)  \in (0.01,0.1)\;.
\end{equation}
Recall that the mass matrix under the FN framework is approximately rank one. If only two generations are present, then one must have $m_2\gg m_1$, which is not desirable. With three generations, one could have $m_3\gg m_2\approx m_1$ under a reasonable choice of $\beta_{ij}$. In the Gemini DM model, we call $\chi_3$ the mother particle and $\chi_{1/2}$ the twins.

\begin{figure}[t]
    \centering
    \includegraphics[width=7.5cm]{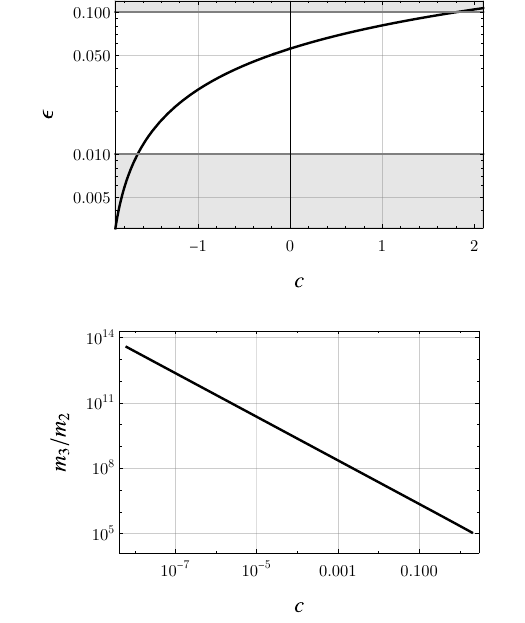}
    \caption{Top: mass difference $\epsilon$ between the twins $\chi_{1/2}$ plotted against the single parameter $c$ describes the $\beta_{ij}$ in Eq.~\eqref{eq:beta}. Bottom: mass ratio between the mother particle and the twins as a function of $c$.
    }
    \label{fig:eigenvalues}
\end{figure}

Here we present a specific parameter choice that gives us a desirable spectrum.
We fix the FN parameter as $\lambda=0.171$~\cite{Qiu:2023igq} as a benchmark model. The value of the FN parameter $\lambda$ has degeneracy with FN charge choices~\cite{Fedele:2020fvh}.
Choose the FN charge as 
\begin{equation}
n_{1}=4.5 + n_3\;,\quad n_{2}=2.5+n_3\;.
\label{eq:FN_charge}
\end{equation}
The charge $n_3$ only determines the overall scale, which is irrelevant when determining couplings and mass ratios.
The half-charge can be absorbed by redefining the FN parameter.
We parametrize $\beta$ as
\begin{equation}
    \beta =\begin{pmatrix}
        1 & 1 & 1+c \\
        1 & 1 & 1\\
        1+c & 1 & 1
    \end{pmatrix}
    \label{eq:beta}
\end{equation}
in order to obtain various spectra by varying the parameter $c$ and determine how to produce the Gemini DM scenario. 
As shown in the top panel of Fig.~\ref{fig:eigenvalues}, for $-1.7\lesssim c \lesssim 2.0$, one achieves the Gemini DM spectrum with an $\epsilon$ that resolves the $S_8$ tension using Eq.~\eqref{eq:Lchi}. Furthermore, the corresponding mass ratio of the much heavier mother particle $\chi_3$ to the Gemini is shown in the bottom panel of Fig.~\ref{fig:eigenvalues}. 

We parametrize the flavon couplings in Eq.~\eqref{eq:Lchi} as
\begin{align}
    g_{ij}^\phi &= \frac{1}{f_a}\left(m_i-m_j + m_i \delta_{ij}\right)  \mathcal{A}_{ij} \nonumber \\
    g_{ij}^a & = \frac{1}{f_a}\left(m_i-m_j + m_i \delta_{ij}\right) \mathcal{B}_{ij}\;,
    \label{eq:gagphi}
\end{align}
where $\mathcal{A}$ and $\mathcal{B}$ are numerical matrices that are calculated explicitly with our choice of $\beta$ [Eq.~\eqref{eq:beta}] and FN charges [Eq.~\eqref{eq:FN_charge}]. By varying $c$, one obtains a one-to-one correspondence between couplings and $\epsilon$. To determine the values and variance of the matrices $\mathcal{A}$ and $\mathcal{B}$ as shown in Table~\ref{tab:AB}, we use the following method. We randomly sample $c$, adopting the uniform distribution as the prior, and select those couplings associated with $\epsilon\in(0.01,0.1)$. From these values we collect the corresponding $\mathcal{A}$ and $\mathcal{B}$. We calculate the average and their deviations, as shown in Table~\ref{tab:AB}. The result here is that the ranges of values in the matrices are not large compared to their central value. We can therefore justifiably treat them as constants for the following calculations.

\begin{table}[t]  \caption{Statistics of matrix elements $|\mathcal{A}_{ij}|^2$ and $|\mathcal{B}_{ij}|^2$ under the parametrization Eqs.~\eqref{eq:FN_charge},~\eqref{eq:beta}, and~\eqref{eq:gagphi} with $\epsilon\in(0.01,0.1)$, and randomly sampled $c$ under a uniform distribution. Central values are averages. The upper and lower uncertainties indicate the maximum and the minimum, respectively.}
    \begin{ruledtabular}
    \begin{tabular}{l c c}
       $(ij)$ &  $|\mathcal{A}_{ij}|^2$ & $|\mathcal{B}_{ij}|^2$  \\
        \midrule
       $(11)$  &  $6.64^{+0.49}_{-0.47} \times 10^3$  & $48.6^{+0.4}_{-0.4}$   \\[0.3em]
       $(22)$ & $6.76^{+0.48}_{-0.48} \times 10^3$ &  $49.4^{+0.3}_{-0.4}$ \\[0.3em]
       $(33)$ & $1.57\times 10^{3} \pm 10^{-3}$ & $12.3 \pm 10^{-5}$ \\[0.3em]
       $(21)$  &  $1.37\times 10^2 \pm 10^{-1}$  & $0.999^{+0.001}_{-0.002}$ \\[0.3em]
       $(31)$ & $5.77^{+0.95}_{-0.88}\times 10^{-2}$ & $4.21^{+0.70}_{-0.65}\times 10^{-4}$ \\[0.3em]
       $(32)$ & $6.79^{+1.11}_{-0.99}\times 10^{-2}$  &  $4.98^{+0.79}_{-0.75}\times 10^{-4}$
    \end{tabular}
    \end{ruledtabular}
    \label{tab:AB}
\end{table}

Since $\phi$ obtains its mass from the FN symmetry breaking, $m_\phi$ is typically associated with the breaking scale $f_a$. The axion $a$ is typically very light, obtaining its mass from some explicit breaking, which we treat as a free parameter and $m_a \ll m_{1/2}$. In line with this thinking, we consider the case that $\chi$ 
dominantly decays through $a$. With $\chi_2$ slightly heavier than $\chi_1$ ($m_2 \gtrsim m_1$), $\chi_3$ decays via $\chi_{1/2}+a$, while $\chi_2$ 
dominantly decays to $\chi_1 + a$. The heavy $\phi$ decays into $\chi$'s and $a$.
These decay widths are given by
\begin{align}
    \Gamma_{\chi_i\to\chi_j a}  & = \frac{m_i |g_{ij}^a|^2}{16\pi}\gamma_+\left(\frac{m_j}{m_i}, \frac{m_a}{m_i}\right) \;,  \nonumber\\
    \Gamma_{\phi \to \chi_k \bar \chi_l} & = \frac{m_\phi |g_{ij}^\phi|^2}{8\pi(1+\delta_{kl})}\gamma_- \left( \frac{m_k}{m_\phi} , \frac{m_l}{m_\phi} \right)\;.
    \label{eq:decaywidth}
\end{align}
Kinematics requires that $m_i>m_j+m_a$ and $m_\phi> m_k+m_l$. The two-body decay phase-space factor $\gamma_\pm$ is
\begin{align*}
    \gamma_\pm(y,z) &= (1+y+z)^{3/2} (1\pm y-z)^{3/2} \\
    &\qquad\qquad \times \sqrt{(1-y+z)(1\mp y-z)}\;,
\end{align*}
for any real number $y$ and $z$. 
There are sub-dominant dark sector decays into SM particles mediated by $\phi$ and $a$, $\chi_i \to \chi_j + {\rm SM} + {\rm SM}$. However, it is a three-body decay, suppressed by at least another coupling squared~$\sim (m_{\rm SM}/f_a)^2$. So the three-body decay branching ratio (Br) is negligible.

\begin{figure}[t]
    \centering
    \includegraphics[width=7.5cm]{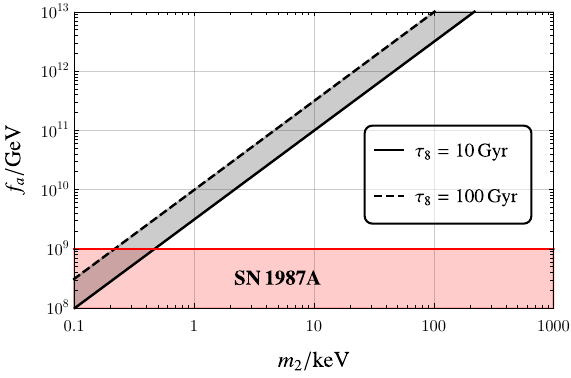}
    \caption{Constraints on the $f_a$ and $m_2$ parameter space. Here we choose $\epsilon=0.05$ and $m_a=10^{-6}~{\rm eV}$ as the benchmark model. The gray stripe is the region that satisfies Eq.~\eqref{eq:Gamma8} and resolves the $S_8$ tension. The red region is excluded by the supernova axions~\cite{Lella:2023bfb}. 
    }
    \label{fig:m2-fa}
\end{figure}

Our twins are DM and the decay $\chi_2 \to \chi_1 + a$ would give the lighter $\chi_1$ enough kinetic energy to wash out some structures, which results in a smaller $S_8$ for the late Universe. As long as the decay width is approximately
\begin{equation}
\Gamma_{\chi_2\to\chi_1 a} = \frac{1}{\tau_8} \sim \mathcal{O}(10^{-43})~{\rm GeV}\;,
\label{eq:Gamma8}
\end{equation}
this decay resolves the $S_8/\sigma_8$ tension and, equivalently, $\tau_8\sim \mathcal{O}(10)$--$\mathcal{O}(100)$~Gyr~\cite{Fuss:2024dam}. It is projected into a region of the $f_a$-$m_{1/2}$ parameter space.\footnote{Note that for $\tau>100\,{\rm Gyr}$, the model still works as a DM model without resolving the $S_8$. However, for $\tau<10\,{\rm Gyr}$, it conflicts with other structure formation~\cite{Fuss:2024dam} constraints.}
Choosing $\epsilon=0.05$ and $m_a=10^{-6}~{\rm eV}$ as a benchmark model, the preferred parameter region could be parametrized as
\begin{equation}
    m_{1/2}\approx 37 \times \left( \frac{f_a^2}{\tau_8}\right)^{1/3}\;,
    \label{eq:m2-fa}
\end{equation}
which is the gray stripe shown in Fig.~\ref{fig:m2-fa}. 

The axion mass $m_a$ puts an upper bound on $f_a$ via its overproduction from the misalignment mechanism. For $m_a\sim 10^{-6}$~eV, this gives $f_a<1.1\times 10^{13}$~GeV~\cite{Hui:2021tkt}.
There is a lower bound on $f_a$ in this figure. Our axion $a$ couples to nucleons with a coupling of order~$\sim 1/f_a$. This means that supernovas can emit a large axion flux, which is constrained as $f_a>10^9$~GeV by the Kamiokande II neutrino detector at the time of SN1987A~\cite{Lella:2023bfb}. 
Furthermore, the axion will couple to the photon via SM fermion loops. This coupling is highly suppressed $g_{a\gamma}\sim m_a^2/f_a^2$, avoids constraints by orders of magnitude~\cite{Adams:2022pbo}, and leaves the axion stable on cosmological time scales.
Note that with $m_2 \sim 100~{\rm keV}$ the three-body $\chi_2$ decay into charged fermions is kinematically forbidden, not just suppressed. However, the $\chi_2\to\chi_1+\gamma+\gamma$ decay occurs through a loop but has a decay width of $\Gamma_{\chi_1 \to\chi_1\gamma\gamma}\lesssim (m_2/f_a)^2\Gamma_{\chi_2\to\chi_1 a}\sim \mathcal{O}(10^{-48})~{\rm s}^{-1}$, well below current limits~\cite{Calore:2022pks}.

The Gemini DM model has several parameters, among which $\epsilon$ fixes the mass ratio between the twins and $\tau_8$ restricts $m_{1/2}$ in terms of the breaking scale $f_a$. For now, the mass of the mother particle $m_3$ and the scale $f_a$ are still undetermined. As long as $m_a$ is small, it does not affect Gemini DM phenomenology. 
In the following section, we show that the production of Gemini DM in the early Universe will allow us to limit the $f_a$ and $m_3$ values that solve the $S_8$ tension. This means that there is essentially only one free parameter, which is $f_a$.

\section{Cosmological production and constraints}\label{sec:production}

We propose a consistent production mechanism for the Gemini DM. 
The mother particle $\chi_3$ 
freezes in through thermal $\phi$ decay and subsequently gives birth to the twins $\chi_{1/2}$ via $\chi_3 \to \chi_{1/2}+a$.
This mechanism could be divided into three stages, which are 1) $\phi$ stays in thermal, 2) $\phi$ decays into $\chi_3$,\footnote{According to Table~\ref{tab:AB} and Eq.~\eqref{eq:gagphi}, $g^\phi_{33}/g^\phi_{22} \approx m_3/m_2 \gg 1$ for the Gemini spectrum we are interested in. Therefore, $\phi$ mainly decays into $\chi_3$.} and 3) $\chi_3$ decays into $\chi_1$ or $\chi_2$.

\subsection{$\phi$ in thermal equilibrium with the Standard Model bath}\label{subsec:phi_thermal}

From Fig.~\ref{fig:m2-fa}, one sees that current experiments require that at least $f_a>10^9~{\rm GeV}$. Meanwhile, the $m_\phi$ is generated from the potential that breaks the $U(1)_{\rm FN}$. This means that $m_\phi$ is much larger than the Higgs mass scale~$\sim \mathcal{O}(100)~{\rm GeV}$. Furthermore, the SM thermal bath temperature required to produce appreciable quantities of $\phi$ is much higher than the electroweak breaking scale.
As reviewed in the Appendix, at this scale, both $\phi$ and $a$ couple to SM particles through
\begin{equation}
n\lambda^{n}\frac{\phi}{f_a} \overline{Q} \mathcal{H} q \;,\quad n\lambda^{n}\frac{a}{f_a} \overline{Q} \mathcal{H} q\;,
\label{eq:4P_interaction}
\end{equation}
where $n=n_Q^{\rm FN}+n_q^{\rm FN}$ is the sum of the quark FN charges.
$\mathcal{H}$ is the SM Higgs doublet that does not carry an FN charge.
We shall see that flavons enter the SM thermal bath through these interactions.
Since we are interested in the temperature, $T\sim\mathcal{O}(m_\phi)$, we treat all SM particles as massless for simplicity.\footnote{For temperatures higher than the vacuum mass of a particle, it is relativistic, which means that its phase-space distribution could be approximated by a massless one. Including the thermal mass $\sim g T$ will bring a small correction to its phase-space distribution, since it is suppressed by the coupling $g$.} Focus on $\phi$ for the moment. There are three different processes associated with this vertex, which are
\begin{align}
    \phi + \mathcal{H} & \to \bar{Q} + q \;, \nonumber\\
    \phi + Q & \to \mathcal{H} + q \;, \nonumber \\
    \phi + q & \to \mathcal{H} + Q \;.
    \label{eq:process}
\end{align}
At the leading order, they are described as having the same amplitude, which is given by
\begin{equation}
|\mathcal{M}|^2  \approx \frac{\lambda^{2}}{f_a^2} (2p_Qp_q)\;,
\end{equation}
where $p_j$ is the four-momentum of the $j$ particle. 
Here we have explicitly chosen $n=1$ because this will be the leading contribution.
The other three particles, namely, $\{\mathcal{H}, Q, q\}$, are in thermal equilibrium.
The equilibrium number density $n_j^{\rm eq}$ can be approximated by integrating the Maxwell-Boltzmann distribution $f_j^{\rm eq}\approx e^{- E_j/T}$,
\begin{align}
    n_j^{\rm eq} & = \frac{g_j}{(2\pi)^3}\int d^3p_j f_j^{\rm eq} \\
    &= \frac{g_j}{2\pi^2}m_j^2 T K_2(m_j/T) \overset{m_j\to 0}{ = } \frac{g_j}{\pi^2} T^3 \nonumber\;.
    \label{eq:neq}
\end{align}
Here $g_j$ is the degree of freedom of the $j$ particle and $m_j$ is its mass. $T$ labels the temperature. 
The last approximation above takes the massless limit, which applies for $\{\mathcal{H}, Q, q\}$. 
$K_\alpha(x)$ is the modified Bessel function of the second kind. 
The number density $n_\phi$ evolution of $\phi$ is governed by the Boltzmann equation, which includes the three processes in Eq.~\eqref{eq:process} and their inverses,
\begin{align}
    \dot{n}_\phi + 3 H n_\phi   &= - \tilde{\Gamma}_\phi \left( n_\phi -  n_\phi^{\rm eq} \right)  \\
    \tilde{\Gamma}_\phi & = \langle \sigma_{\mathcal{H}} v \rangle n_{\mathcal{H}}^{\rm eq} +  \langle \sigma_Q v \rangle n_Q^{\rm eq} +\langle \sigma_q v \rangle n_q^{\rm eq} \;, \nonumber 
    \label{eq:thermal_xsec_tot}
\end{align}
where $\langle \sigma_j v \rangle$ labels the thermally averaged cross section of the process whose initial states are $\phi$ and $j$ particles.
$H(T)$ is the Hubble parameter that describes the expansion of the Universe.
The interaction rate $\tilde{\Gamma}_\phi(T)$ and the Hubble expansion rate $3H(T)$ determine whether the $\phi$ stays in thermal equilibrium.
Since at the relevant temperatures the particles $\{\mathcal{H}, Q, q\}$ are all massless, one can simplify the expression for $\tilde{\Gamma}_\phi$ by performing a relabeling of the integration variables in the second and third terms in $\tilde{\Gamma}_\phi$. This allows us to write
\begin{align}
    \tilde{\Gamma}_\phi & \approx \frac{1}{n_\phi^{\rm eq}} \int \prod_j d\Pi_j \frac{\lambda^{2}}{f_a^2}\left( 2p_Qp_q + 2p_{\mathcal{H}} p_q + 2p_Q p_{\mathcal{H}}\right) \\
    &\qquad \times (2\pi)^4 \delta^4(p_\phi+p_{\mathcal{H}} - p_Q -p_q) e^{-(E_\phi + E_\mathcal{H})/T} \;,\nonumber
\end{align}
where the phase-space integral is defined as $d\Pi_j = g_j d^3p_j/[(2\pi)^3(2E_j)]$ and $j$ runs for all four particles in these processes. Following the method in Ref.~\cite{Gondolo:1990dk}, one converts the above formula into a single integration,
\begin{align}
  \tilde{\Gamma}_\phi 
& \approx 
  \frac{g_{\mathcal{H}} g_Q g_q \lambda^{2}}{ 16 (2\pi)^3}
  \frac{ T^5}{f_a^2 m_\phi^2} 
  \frac{\mathcal{I}(m_\phi/T)}{K_2(m_\phi/T)} 
\;, \label{eq:phi_int_rate}\\ 
    \mathcal{I}\left(\zeta\right) 
& = 
    \int_{\zeta}^\infty d\xi \left(\xi^2 - \zeta^2\right) \left(2\xi^2 - \zeta^2 \right)K_1(\xi)\;, \nonumber
\end{align}
where $\zeta$ is a real number.
Taking the standard cosmology picture, we assume that the Universe is dominated by radiation at early times. Then, the Hubble parameter can be approximated by 
\begin{equation}
    H(T)\approx\frac{\pi}{3}\sqrt{\frac{g_\star}{10}}\frac{T^2}{M_{\rm Pl}}\approx 3.4 \times \frac{T^2}{M_{\rm Pl}}\;,
\end{equation}
where $M_{\rm Pl}=2.4\times 10^{18}~{\rm GeV}$ is the reduced Planck scale and $g_\star$ is the temperature-dependent effective number of energy degrees of freedom. To get the numerical expression, we take $g_\star(T\gtrsim T_{\rm EWSB})\approx 107$~\cite{Drees:2015exa}.

\begin{figure}[t]
    \centering
    \includegraphics[width=7.5cm]{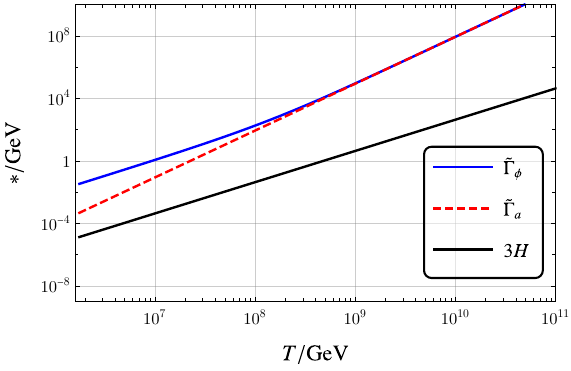}
    \caption{Here we choose $f_a =2\times 10^{10}$ GeV and $m_\phi=10^9$ GeV. The degrees of freedom are $g_Q=12$, $g_q=6$, and $g_{\mathcal{H}}=4$. The FN charge is chosen as $n=1$ as the leading contribution, and the FN parameter is $\lambda=0.171$. Note that the red-dashed line, the interaction rate for the axion $\tilde{\Gamma}_a$, crosses with the Hubble parameter at $T_{\rm dec}^a \approx 5 \times 10^4$ GeV under this choice of parameters.}
    \label{fig:gamma-H}
\end{figure}

To estimate when $\phi$ decouples from the SM bath, we compare $\tilde{\Gamma}_\phi(T)$ and $3H(T)$. Similarly, this can be done for the $a$ particle, and the interaction rate of~\eqref{eq:phi_int_rate} is modified by taking the massless limit
\begin{equation}
    \tilde{\Gamma}_a \approx \frac{g_{\mathcal{H}} g_Q g_q \lambda^{2}}{(2 \pi)^3} \frac{ T^3}{f_a^2}\;.
    \label{eq:Gamma_a}
\end{equation}
In Fig.~\ref{fig:gamma-H} we show a specific realization where $\tilde{\Gamma}_\phi>3H$ for temperatures below the breaking scale $f_a$. We see that the axion interaction rate $\tilde{\Gamma}_a$ follows the simple $T^3$ dependence, which decreases in a cooling Universe faster than Hubble. This leads to an axion decoupling at around $\sim 10^5~{\rm GeV}$ (left side of figure)~\cite{Lillard:2018zts}. 
Interestingly, for massive $\phi$ the temperature dependence changes below $T\sim m_{\phi}$, keeping it in equilibrium, and its number density can be described by $n_\phi^{\rm eq}$. 

\subsection{$\chi_3$ freeze-in from $\phi$ decay}\label{sec:FI_solving_Boltz}

Given that there are parameter choices for our model that ensure that $\phi$ stays in the thermal bath, we consider the standard freeze-in for $\chi_3$ from $\phi$ decay~\cite{McDonald:2001vt,Kusenko:2006rh,Petraki:2007gq,Hall:2009bx,Bernal:2017kxu,Bauer:2017qwy}.
The Boltzmann equation for $\chi_3$ is 
\begin{equation}
    \dot{n}_{\chi_3} + 3H n_{\chi_3} = S(\phi \to \chi_3 \bar \chi_3) \;,
    \label{eq:boltzmann_chi_3}
\end{equation}
where the source term is given by
\begin{align}
    S(\phi \to \chi_3 \bar \chi_3)  & = g_\phi g_3^2 \int \frac{d^3 p_\phi}{(2\pi)^3} \frac{m_\phi}{E_\phi} \Gamma_{\phi \to \chi_3 \bar \chi_3} e^{-E_\phi/T} \nonumber \\
    & = \frac{g_\phi g_3^2}{2\pi^2}  \Gamma_{\phi \to \chi_3 \bar \chi_3} m_\phi^2 T K_1(m_\phi/T)\;.
\end{align}
The decay width $\Gamma_{\phi \to \chi_3 \bar \chi_3}$ is given by Eq.~\eqref{eq:decaywidth}. 
In the second line, we have evaluated the integration explicitly.
Following standard parametrization, we define the yield $Y_3= n_{\chi_3}/s$, where $s$ is the entropy density. The equilibrium condition of conservation of total entropy implies that $ds/dt=-3 s H$. So the lhs of the Boltzmann equation can be written as $ \dot{n}_{\chi_3} + 3 H n_{\chi_3}= s dY_3/dt$.
Furthermore, defining the variable $x= m_\phi/T$ enables one to write $dx/dt = -(x/T)(dT/dt) = xH$ for the radiation-dominated era.
Then, the Hubble parameter can be written as $H(x) \approx 3.4\times  m_\phi^2/M_{\rm Pl} x^2$.
Similarly, we extract out the $x$ dependence in the entropy density, $s \approx 27.3\times  s_0 m_\phi^3 /T_0^3 x^3$,
where $T_0=2.3\times 10^{-13}$ GeV is the cosmic microwave background (CMB) radiation temperature today.
Therefore, the Boltzmann equation for $\chi_3$ [Eq.~\eqref{eq:boltzmann_chi_3}] becomes
\begin{equation}
    \frac{dY_3}{dx} \approx \frac{g_\phi g_3^2}{186 \pi^2} \frac{ T_0^3 M_{\rm Pl}}{s_0 m_\phi^2}  \Gamma_{\phi \to \chi_3 \bar \chi_3}  x^3 K_1(x)\;,
\end{equation}
which can be integrated directly, either numerically or by taking $\int_0^\infty x^3K_1(x) = 3\pi/2$.
If we assume that the initial yield of $\chi_3$ is zero, we obtain the freeze-in yield of $\chi_3$,
\begin{align}
    Y_3^{\rm f.i.}  & \approx Y_3(\infty)  \approx\frac{3g_\phi g_3^2}{371
    \pi}\frac{ T_0^3 M_{\rm Pl}}{s_0 m_\phi^2}  \Gamma_{\phi \to \chi_3 \bar \chi_3} \\ 
    & \approx 0.32 \times \frac{T_0^3 M_{\rm Pl}}{s_0 m_\phi}  \frac{m_3^2}{f_a^2} \left(1- \frac{4m_3^2}{m_\phi^2} \right)^{3/2} \nonumber\;.
\end{align}
In the last line, we used $g_\phi = 1$ and $g_3 = 2$ and took the numerical value for $|\mathcal{A}_{33}|^2$ in the decay width from Table~\ref{tab:AB}.

\subsection{Relic $\chi_{1/2}$ from $\chi_3$ decay}

\begin{figure}[t]
    \centering
    \includegraphics[width=7.5cm]{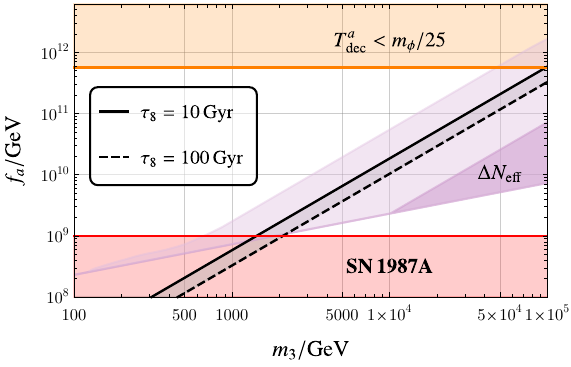}
    \caption{Here we choose $m_\phi=10^{9}$ GeV and $\Omega_{\rm DM}h^2=0.12$ under the benchmark model $\epsilon=0.05$ and $m_a=10^{-6}$ eV. The gray stripe is the parameter range that gives the correct relic and resolves $S_8$. The orange region is excluded by $T_{\rm dec}^a>m_\phi/25$ and the red region is excluded by the supernova axions~\cite{Lella:2023bfb}. The darker purple region is excluded by overproduction of the dark radiation from $\Delta N_{\rm eff}<0.276$~\cite{Planck:2018vyg}. The lighter-purple region indicates $\Delta N_{\rm eff}>0.04$ that can be covered for the future CMB-S4~\cite{CMB-S4:2016ple}, whose sensitivity could reach $\Delta N_{\rm eff} \sim 0.02$.
    }
    \label{fig:relic_fa-m3}
\end{figure}

In the Gemini DM model, DM consists of two twins, namely, $\chi_1$ and $\chi_2$. They are produced from $\chi_3$ decay via $\chi_3 \to \chi_{1/2} + a$.
One can write the relic yield of $\chi_i$ as $Y_i \approx {\rm Br}(\chi_3\to \chi_i a)Y_3^{\rm f.i.}$\footnote{The branching ratio can be approximated by using values for $|\mathcal{B}_{ij}|^2$ in Table~\ref{tab:AB}. For example, ${\rm Br}(\chi_3 \to \chi_2 a) \approx |\mathcal{B}_{32}|^2/(|\mathcal{B}_{32}|^2 + |\mathcal{B}_{31}|^2)$.} and we have $Y_1 + Y_2 \approx Y_3^{\rm f.i.}$.
The energy density today is given by $\rho_i(T_0) = m_{1/2} Y_i s_0$, where we have approximated $m_1\approx m_2 \approx m_{1/2}$.
Now the total relative relic density of the Gemini DM is given by
\begin{align}
  \Omega_{\rm DM} h^2
& =
 \frac{(\rho_1 + \rho_2) h^2}{ 3 M_{\rm Pl}^2 H_0^2 }
\approx 
 \frac{m_{1/2} Y_3^{\rm f.i.} s_0 h^2}{ 3 M_{\rm Pl}^2 H_0^2 }
\label{eq:relic}\\
& \approx
  0.11 \times \frac{h^2 T_0^3 m_{1/2}}{ H_0^2 M_{\rm Pl}  m_\phi} \frac{m_3^2}{f_a^2} \left( 1-\frac{4m_3^2}{m_\phi^2}\right)^{3/2} \;.\nonumber
\end{align}
To resolve the $S_8$ tension, the decay width of $\Gamma_{\chi_2 \to \chi_1 a}$ satisfying Eq.~\eqref{eq:Gamma8} puts a relation between $m_{1/2}$ and $f_a$. For the benchmark model with $\epsilon=0.05$ and $m_a=10^{-6}$ eV~\eqref{eq:m2-fa}, this gives
\begin{align}
  \Omega_{\rm DM} h^2 
& \approx 
  4.0 \times  
  \frac{h^2T_0^3 m_3^2}{ H_0^2  M_{\rm Pl} m_\phi (\tau_8 f_a^4)^{1/3}} 
  \left( 1-\frac{4m_3^2}{m_\phi^2}\right)^{3/2} \\
& \approx 
  0.12 
  \left(\frac{m_3}{ 1.1 \times 10^4~{\rm GeV}}\right)^2 
  \left(\frac{f_a}{2\times 10^{10}~{\rm GeV}}\right)^{-4/3} \nonumber\;,
\label{eq:Relic_DM_fi}
\end{align}
where the Hubble parameter today is $H_0/h \approx 2.1\times 10^{-42}~{\rm GeV}$ (from $H_0 \equiv 100 h~{\rm km/s/Mpc} $)~\cite{ParticleDataGroup:2022pth}.
For the last line, we have chosen $\tau_8 = 10~{\rm Gyr}$, and $m_\phi = 10^{9}~{\rm GeV}$. 
To give the correct relic that fits the cosmology, we should have $\Omega_{\rm DM}h^2\approx 0.12$.
Figure~\ref{fig:relic_fa-m3} shows the allowed parameter space for $f_a$ and $m_3$ that resolves $S_8$ and indicates the correct Gemini DM relic by the gray shaded strip.
For a fixed $m_\phi=10^{9}$~GeV, we have a preferred $f_a\sim \mathcal{O}(10^{9})$--$\mathcal{O}(10^{12})$~GeV, which indicates that the Gemini DM mass should be roughly $m_{1/2}\sim \mathcal{O}(1)$--$\mathcal{O}(100)$~keV, according to Fig.~\ref{fig:m2-fa}.

In addition to potential probes of this model from structure formation, the fact that the 
axion was once in thermal equilibrium means that axion relics will potentially be observable today. References~\cite{Turner:1986tb,Salvio:2013iaa,DEramo:2021lgb,Bouzoud:2024bom} and others explored this possibility for the QCD axion. The greatest contribution will be by altering the number of relativistic degrees of freedom $N_{\rm eff}$ in the form of dark radiation. Deviations from the SM prediction $(N_{\rm eff})_{\rm SM}\approx 3.044$~\cite{Mangano:2001iu,deSalas:2016ztq,Gariazzo:2019gyi,Akita:2020szl,Froustey:2020mcq,Bennett:2020zkv, Drewes:2024wbw} are parametrized by 
\begin{equation}
\Delta N_{\text{eff}}= \left.\frac{8}{7}\left(\frac{11}{4}\right)^{4/3}\frac{\rho_a}{\rho_\gamma}\right\vert_{T_{\rm rec}}\;,
\label{eq:delta_N_eff}
\end{equation}
where $\rho_\gamma$ is the photon energy density and $T_{\rm rec}\approx 0.26$ eV is the recombination temperature~\cite{Rubakov:2017xzr}.
The energy density ratio is evaluated at the time of the recombination because measurements of the CMB give the most stringent constraints of $\Delta N_{\rm eff}$. For example, the {\it Planck} Collaboration gives $(N_{\rm eff})_{\rm P18}=2.88^{+0.44}_{-0.42}$~\cite{Planck:2018vyg}, which leads to $\Delta N_{\rm eff}=(N_{\rm eff})_{\rm P18}-(N_{\rm eff})_{\rm SM}\leq 0.276$. The energy density of the thermal axion at decoupling is given by $\rho_a^{\rm th}(T_{\rm dec}^a)= \rho_\gamma(T_{\rm dec}^a)/2=\pi^2(T_{\rm dec}^a)^4/30$, where $T_{\rm dec}^a$ can be estimated by evaluating $3H\approx\tilde{\Gamma}_a$. 
Using Eq.~\eqref{eq:Gamma_a}, one obtains $T_{\rm dec}^a\approx 300 f_a^2/M_{\rm Pl}$. 
The thermal axion's contribution to the $\Delta N_{\rm eff}$ is proportional to
\begin{equation}
\left.\frac{\rho_a^{\rm th}}{\rho_\gamma}\right|_{T_{\rm rec}} = \frac{1}{2} \left(\frac{g_{\star S}(T_{\rm rec})}{g_{\star S}(T_{\rm dec}^a)}\right)^{4/3} \;,
    \label{eq:rho_a_th}
\end{equation}
where $g_{\star S}(T)$ is the relativistic degrees of freedom in entropy.
This contribution has $f_a$ dependence through $T_{\rm dec}^a$.
In the Gemini DM model, we require $f_a>10^9$ GeV (see Fig.~\ref{fig:m2-fa}), which translates to $\Delta N_{\rm eff}\approx 0.028$. As $f_a$ increases, the decoupling temperature $T_{\rm dec}^a$ increases, which indicates a larger $g_{\star S}(T_{\rm dec}^a)$ and a smaller $\Delta N_{\rm eff}$, with a minimum value of $\approx 0.027$ assuming no extra beyond-the-SM degrees of freedom in thermal equilibrium at early times. 
This constitutes a falsifiable prediction for the Gemini DM model because next-generation CMB probs are expected to reach a sensitivity at or below this~\cite{CMB-S4:2016ple,CMB-HD:2022bsz}.
Furthermore, there are two possible additional ways in which the axion energy density can be enhanced to levels that are already ruled out. 

\begin{enumerate}[label=(\roman*),leftmargin=15pt]
\item   In addition to the decay that produces $\chi_3$, the $\phi$ particle also decays into the axion via $\phi\to a+ a$. The decay term is generated from the kinetic term of $\Phi$ and is therefore likely to dominate.
As long as this decay occurs when $a$ is still in thermal equilibrium, the effect will be washed out. The production from $\phi$ decay continuously happens until $n_\phi$ is exponentially suppressed when $T \lesssim m_\phi/25$. Therefore, as long as $T_{\rm dec}^a< m_\phi/25$, there is no appreciable additional production of the dark radiation from $\phi$ decay. 
Therefore, we avoid the overproduction of the dark radiation from $\phi$ decay by putting an upper bound for $f_a$, indicated by the orange exclusion region in Fig~\ref{fig:relic_fa-m3}.

\item The decay $\chi_3 \to \chi_{1/2} + a$ can also contribute to dark radiation. It occurs roughly at the temperature $3H(T_3) \approx \Gamma_{\chi_3 \to \chi_2 a}$, which is 
\begin{equation}
    T_3
\approx 
  9.9 \times 10^{-4} 
  \frac{\sqrt{m_3^3M_{\rm Pl}}}{f_a}\;.
\label{eq:T_3}
\end{equation}
For the parameter space that we are interested in, we usually have $T_3 < T_{\rm dec}^a$, so this extra contribution could not be avoided.
To estimate the additional contribution we assume instantaneous decay, i.e., $\Delta\rho_a(T_3) \approx m_3 Y_3^{\rm f.i.}s(T_3)/2$, where the half indicates the energy sharing between $\chi_{1/2}$ and $a$. This gives an additional contribution to the $\Delta N_{\rm eff}$ at the recombination, which is proportional to
\begin{align}
    \left.\frac{\Delta \rho_a}{\rho_\gamma} \right|_{T_{\rm rec}} & = \frac{\Delta \rho_a(T_3)}{\rho_\gamma(T_{\rm rec})}\left(\frac{T_{\rm rec}}{T_3}\right)^4\left(\frac{g_{\star S}(T_{\rm rec})}{g_{\star S}(T_3)}\right)^{4/3} \nonumber\\
    & \approx 6.9\times 10^6 \frac{T_3}{m_\phi}\left(\frac{g_{\star S}(T_{\rm rec})}{g_{\star S}(T_3)}\right)^{4/3}\;, \label{eq:delta_rho_a}
\end{align}
where we have neglected the phase space factor from Eq.~\eqref{eq:decaywidth} in the second line since we are mostly interested in the limit $m_\phi \gg m_3$. This energy density ratio depends on $m_3$ and $f_a$ through $T_3$ given in Eq.~\eqref{eq:T_3}. The constraint demands an upper bound of this ratio, which is translated to a lower bound on $f_a$ in terms of $m_3$.
\end{enumerate}
Therefore, the total radiation $\rho_a = \rho_a^{\rm th} + \Delta \rho_a$ at recombination gives an $\Delta N_{\rm eff}$ that is proportional to the sum of Eqs.~\eqref{eq:rho_a_th} and~\eqref{eq:delta_rho_a}.
We show a region in Fig.~\ref{fig:relic_fa-m3} that is excluded by current CMB results on $\Delta N_{\rm eff}>0.276$ reported by {\it Planck}~\cite{Planck:2018vyg}. 
Once again, the above calculation requires $T_{\rm dec}^a > T_3$, otherwise, the decay axions are thermalized and $\Delta N_{\rm eff}\approx 0.027$. This is the reason why the low-$f_a$ region is not covered by the $\Delta N_{\rm eff}$ constraint in Fig.~\ref{fig:relic_fa-m3}. 
We have additionally shaded a region in light purple that shows where $\Delta N_{\rm eff}>0.04$. We do this to show the level of sensitivity future CMB probes will require in order to rule out this solution to the $S_8$ tension.

Dark matter models produced from thermal processes with $m_{\rm DM}\sim \mathcal{O}(10)$ keV are typically highly constrained by inferring the matter power spectrum through measurements of the Lyman-$\alpha$ forest~\cite{Viel:2005qj,Boyarsky:2008xj}. 
To see whether the WDM constraint applies, one has to estimate the free-streaming scale.
In the instantaneous decay approximation of the process $\chi_3\to\chi_{1/2}+a$, the twins $\chi_{1/2}$ obtain 
the average momentum 
$ \langle p \rangle_3  \approx m_3/ 2$ at temperature $T_3$ [Eq.~\eqref{eq:T_3}].
This momentum is redshifted to today and becomes $\langle p \rangle_0$, rendering the twins nonrelativistic. Their 
average velocity today can be expressed as
\begin{align}
  \langle v \rangle_0
&\approx \frac{\langle p\rangle_0}{m_{1/2}} \approx
  \frac{ m_3 }{2 m_{1/2}}
  \frac{T_0}{T_3}
  \left( \frac{g_{\star S}(T_0)}{g_{\star S}(T_3)} \right)^{1/3}\\
&\approx 
  1.4\times 10^{-6} \left(\frac{m_\phi}{10^9\,{\rm GeV}}\right)^{-1/4}\;. \nonumber
\end{align}
For the last approximation, we have evaluated the number under the benchmark model applying Eqs.~\eqref{eq:m2-fa} and Eq.~\eqref{eq:relic} with fixed $\Omega_{\rm DM}h^2=0.12$ and $\tau_8=10$ Gyr, which corresponds to the parameters located in the gray strip in Fig.~\ref{fig:relic_fa-m3}. The dependence on $m_3$ and $f_a$ cancels out after assuming that $m_3\ll m_\phi$. The dependence on $m_\phi$ is weak.
This indicates that the Gemini DM is typically 
very cold today.
With this result, we can estimate the free-streaming scale
of the Gemini DM, which is
\begin{equation}
\lambda_{\rm fs}\approx \frac{ \langle v \rangle_0}{H_0} \approx 4.1 \times 10^{-3}~{\rm Mpc}/h\;.
\label{eq:lambdafs}
\end{equation}
This free-streaming scale 
is much smaller than $\mathcal{O}(1)$ Mpc,
the scale of the Ly-$\alpha$ matter power spectrum~\cite{Murgia:2017lwo}. 
Hence, the usual keV WDM constraint~\cite{Drewes:2016upu} does not apply to Gemini DM~\cite{Garny:2018ali,Decant:2021mhj}.

\section{Summary and discussion}\label{sec:summary}

In this work, we constructed a decaying DM model, named Gemini DM, that resolves the $S_8/\sigma_8$ tension in the large-scale structure, by extending a consistent FN symmetry, which explains the fermion mass hierarchy, to the dark sector.
Our model consists of three generations of dark fermions, among which two have almost degenerate mass (the twins) and the other one (the mother) is much heavier.
The twins are the main DM component today and the mother particle gives birth to them through its decay. 
The production mechanism of the mother particle is nonthermal freeze-in from the SM thermal bath.
The mass of the twins is roughly $m_{1/2} \sim \mathcal{O}({\rm keV})$, which is small. However, because they never thermalize and are produced via $\chi_3$ decay, they constitute a cold DM model.
Dark radiation is also produced during the production of the Gemini DM, which is the prediction of our model. This radiation is expected be probed by future CMB observations.

Unlike the usual DDM solution to the $S_8$ tension, only a fraction of the total DM decays. 
This will likely alter the required $\epsilon\in (0.01,0.1)$, but not by much, for the following reason.
The cold DM decays into warm DM and suppresses structure formation below the free-streaming scale $k_{\rm fs}^{-1}$.
This suppression follows the relation 
$\delta_{\rm WDM} \approx ( k_{\rm fs}^2 /k^2 ) \delta_{\rm CDM}$~\cite{Ringwald:2004np},
where the free-streaming scale is proportional to the particle velocity, determined by the mass separation,
{\it i.e.}, $k_{\rm fs}^{-1} \propto \epsilon$.
If WDM is produced from only a fraction of the total DM,
the suppression of structure formation is
$f_{\rm W} \delta_{\rm WDM} 
\propto f_{\rm W} k_{\rm fs}^2 
\propto f_{\rm W} / \epsilon^2$,
where $f_{\rm W}$ is the fraction of WDM relative to the total DM.
This implies that a smaller fraction of DDM
requires a larger energy release to achieve the same level of structure suppression.
Therefore, the required mass-splitting factor is $\epsilon' \equiv \epsilon / \sqrt{f_{\rm W}}$.
For Gemini DM, where $f_{\rm W} \approx 0.5$, 
this requires $\sqrt{2}$ times the energy 
compared to the previously favored $\epsilon$ in~\cite{Fuss:2024dam}.
Hence, adopting $\epsilon = 0.05$ for our benchmark model in the main text is reasonable.

One may attempt to identify the Gemini DM as the sterile neutrino. This could be achieved by considering the additional coupling between leptons and $\chi$ particles. After the seesaw mechanism, one ends up with a suppressed neutrino mass matrix and a heavy dark sector. We leave this for future study.

If the $S_8$ tension persists and the next generation of CMB measurements find an indication of additional relativistic degrees of freedom $\Delta N_{\rm eff}\approx 0.027$, we will have a strong candidate solution to both simultaneously in the Gemini DM model.

\begin{acknowledgements}
We thank Lorenzo Calibbi, Yong Du, Jie Sheng, Chuan-Yang Xing, and Jiang Zhu for their valuable discussions. The authors are supported by the National Natural Science Foundation of China (12375101, 12090060 and 12090064) and the SJTU Double First Class start-up fund (WF220442604).
\end{acknowledgements}

\appendix

\section{Review of the FN mechanism}\label{appendix:FN}

The FN mechanism explains the SM fermion hierarchy via the introduction of an extra chiral $U(1)_{\rm FN}$ symmetry under some energy scale, which could be either global or gauged. For simplicity, we adopt the global symmetry here. We follow Ref.~\cite{Qiu:2023igq}, and one typical $U(1)_{\rm FN}$ charge assignment of the SM fermions is presented in Table~\ref{tab:FN_charges}. 
Since FN charges are assigned for SM particles,
the usual Yukawa interactions are forbidden by $U(1)_{\rm FN}$.
The leading-order terms that give fermion mass are nonrenormalizable. These nonrenormalizable operators generate interactions between fermions and flavons.  
\begin{table}[t]
    \caption{FN charge assignment of quarks and leptons.}
    \begin{ruledtabular}
    \begin{tabular}{c c c c c c c}
        Generation $i$ &  $\overline{Q}_{\rm L}$ & $u_{\rm R}$ & $d_{\rm R}$ & $\overline{\ell}_{\rm L}$ & $e_{\rm R}$ &  $\Phi$ \\
        \midrule
       $1$  &  $3$  &  $4$  & $1$ & $1$ & $4$ & \multirow{3}{*}{-1}\\
       $2$  & $2$ & $1.5$ & $0$  & $0.5$ & $1$\\
       $3$  & $0$ &  $0$ & $0$ & $0$ & $0$ 
    \end{tabular}
    \end{ruledtabular}
    \label{tab:FN_charges}
\end{table}
We review the FN framework by describing how it works in the lepton sector, but the quark sector can be derived in the same way.
Below some UV cutoff scale $\Lambda$, the leading-order operators are
\begin{equation}
   - \mathcal{L} \supset  g_{ij} \left(\frac{\Phi}{\Lambda}\right)^{n_\ell^{ij}}\overline{\ell}_{\rm L}^i \mathcal{H} e_{\rm R}^j + {\rm H.c.}\;,
   \label{eq:lepton_FN}
\end{equation}
where $\mathcal{H}$ is the Higgs doublet and it does not carry an FN charge. $g_{ij}$ is some order $1$ coupling that is not responsible for the hierarchy. $n_\ell^{ij} = n_\ell^i + n_e^j$ is the lepton FN charge matrix, where $n_\ell^i$ and $n_e^j$ are FN charges of $\overline{\ell}_{\rm L}^i$ and $e_{\rm R}^j$ respectively. Explicitly,
\begin{equation}
    n_\ell = 
    \begin{pmatrix}
        5 & 2 & 1 \\
        4.5 & 1.5 & 0.5 \\
        4 & 1 & 0 \\
    \end{pmatrix}\;.
\end{equation}
Note that here the half charge could be absorbed by redefining the FN parameter.
After the spontaneous symmetry breaking of $U(1)_{\rm FN}$ by $\Phi$, we can do the same field expansion as in Eq.~\eqref{eq:expansion}. After performing the same phase rotation as described in Sec.~\ref{sec:model},
the leading term in Eq.~\eqref{eq:lepton_FN} would be the usual Yukawa interactions 
The Yukawa matrix of the charged leptons is now hierarchically textured,
\begin{equation}
    y_\ell \sim 
    \begin{pmatrix}
        \mathcal{O}(\lambda^5) & \mathcal{O}(\lambda^2) & \mathcal{O}(\lambda^1) \\
        \mathcal{O}(\lambda^{4.5}) & \mathcal{O}(\lambda^{1.5}) & \mathcal{O}(\lambda^{0.5}) \\
        \mathcal{O}(\lambda^4) & \mathcal{O}(\lambda^1) & \mathcal{O}(1)
    \end{pmatrix}\;,
\end{equation}
where $\lambda\approx 0.171$ could give rise to a hierarchy between three generations and provide the best fit with the observation~\cite{Qiu:2023igq}. 
Therefore, after the electroweak symmetry breaking $\mathcal{H} \to \langle \mathcal{H} \rangle$, one obtains lepton masses and interactions with the flavon field,
\begin{equation}
    -\mathcal{L} \supset m_{k} \bar e_{\rm L}^k e_{\rm R}^k +  g_{e,ij}^\phi \phi \bar e_{\rm L}^i e_{\rm R}^j + g_{e,ij}^a a \bar e_{\rm L}^i e_{\rm R}^j + {\rm H.c.}\;
\end{equation}
Here we have rotated in the lepton mass eigenstates.
In this basis, one obtains the relation 
\[
g_{e,ij} \propto \frac{m_i-m_j + m_i \delta_{ij}}{f_a}\;,
\]
which is characteristic of the FN framework.
The diagonal couplings between the flavon and fermions are proportional to the fermion mass, and the off-diagonal couplings are determined by their mass difference. 
The quarks couple to flavons in basically the same way as the charged leptons, with up to slightly different constants. Note that neutrino mass could be generated through a five-dimensional operator~\cite{Weinberg:1979sa,Yanagida:1979as} in the SM effective field theory. This operator should also be modified under the FN framework~\cite{Qiu:2023igq}, which gives rise to neutrino masses and interactions with flavons in a similar manner.

\bibliographystyle{utphys}
\bibliography{Gemini_DM_draft}

\providecommand{\href}[2]{#2}\begingroup\raggedright\begin{thebibliography}{10}

\bibitem{Planck:2018vyg}
{\bfseries Planck} Collaboration, N.~Aghanim {\em et~al.}, ``{\it {Planck 2018
  results. VI. Cosmological parameters}},''
  \href{http://dx.doi.org/10.1051/0004-6361/201833910}{{\em Astron. Astrophys.}
  {\bfseries 641} (2020) A6},
  [\href{http://arxiv.org/abs/1807.06209}{{\ttfamily arXiv:1807.06209}}
  [astro-ph.CO]]. [Erratum: Astron.Astrophys. 652, C4 (2021)].

\bibitem{Abdalla:2022yfr}
E.~Abdalla {\em et~al.}, ``{\it {Cosmology intertwined: A review of the
  particle physics, astrophysics, and cosmology associated with the
  cosmological tensions and anomalies}},''
  \href{http://dx.doi.org/10.1016/j.jheap.2022.04.002}{{\em JHEAp} {\bfseries
  34} (2022) 49--211}, [\href{http://arxiv.org/abs/2203.06142}{{\ttfamily
  arXiv:2203.06142}} [astro-ph.CO]].

\bibitem{DiValentino:2020vvd}
E.~Di~Valentino {\em et~al.}, ``{\it {Cosmology Intertwined III: $f \sigma_8$
  and $S_8$}},''
  \href{http://dx.doi.org/10.1016/j.astropartphys.2021.102604}{{\em Astropart.
  Phys.} {\bfseries 131} (2021) 102604},
  [\href{http://arxiv.org/abs/2008.11285}{{\ttfamily arXiv:2008.11285}}
  [astro-ph.CO]].

\bibitem{Turner:1984ff}
M.~S. Turner, ``{\it {Cosmology with Decaying Particles}},''
  \href{http://dx.doi.org/10.1103/PhysRevD.31.1212}{{\em Phys. Rev. D}
  {\bfseries 31} (1985) 1212}.

\bibitem{Aoyama:2014tga}
S.~Aoyama, T.~Sekiguchi, K.~Ichiki, and N.~Sugiyama, ``{\it {Evolution of
  perturbations and cosmological constraints in decaying dark matter models
  with arbitrary decay mass products}},''
  \href{http://dx.doi.org/10.1088/1475-7516/2014/07/021}{{\em JCAP} {\bfseries
  07} (2014) 021}, [\href{http://arxiv.org/abs/1402.2972}{{\ttfamily
  arXiv:1402.2972}} [astro-ph.CO]].

\bibitem{Enqvist:2015ara}
K.~Enqvist, S.~Nadathur, T.~Sekiguchi, and T.~Takahashi, ``{\it {Decaying dark
  matter and the tension in $\sigma_8$}},''
  \href{http://dx.doi.org/10.1088/1475-7516/2015/09/067}{{\em JCAP} {\bfseries
  09} (2015) 067}, [\href{http://arxiv.org/abs/1505.05511}{{\ttfamily
  arXiv:1505.05511}} [astro-ph.CO]].

\bibitem{FrancoAbellan:2020xnr}
G.~Franco~Abell\'an, R.~Murgia, V.~Poulin, and J.~Lavalle, ``{\it {Implications
  of the $S_8$ tension for decaying dark matter with warm decay products}},''
  \href{http://dx.doi.org/10.1103/PhysRevD.105.063525}{{\em Phys. Rev. D}
  {\bfseries 105} no.~6, (2022) 063525},
  [\href{http://arxiv.org/abs/2008.09615}{{\ttfamily arXiv:2008.09615}}
  [astro-ph.CO]].

\bibitem{FrancoAbellan:2021sxk}
G.~Franco~Abell\'an, R.~Murgia, and V.~Poulin, ``{\it {Linear cosmological
  constraints on two-body decaying dark matter scenarios and the S8
  tension}},'' \href{http://dx.doi.org/10.1103/PhysRevD.104.123533}{{\em Phys.
  Rev. D} {\bfseries 104} no.~12, (2021) 123533},
  [\href{http://arxiv.org/abs/2102.12498}{{\ttfamily arXiv:2102.12498}}
  [astro-ph.CO]].

\bibitem{Simon:2022ftd}
T.~Simon, G.~Franco~Abell\'an, P.~Du, V.~Poulin, and Y.~Tsai, ``{\it
  {Constraining decaying dark matter with BOSS data and the effective field
  theory of large-scale structures}},''
  \href{http://dx.doi.org/10.1103/PhysRevD.106.023516}{{\em Phys. Rev. D}
  {\bfseries 106} no.~2, (2022) 023516},
  [\href{http://arxiv.org/abs/2203.07440}{{\ttfamily arXiv:2203.07440}}
  [astro-ph.CO]].

\bibitem{Fuss:2022zyt}
L.~Fu\ss{} and M.~Garny, ``{\it {Decaying Dark Matter and
  Lyman-\ensuremath{\alpha} forest constraints}},''
  \href{http://dx.doi.org/10.1088/1475-7516/2023/10/020}{{\em JCAP} {\bfseries
  10} (2023) 020}, [\href{http://arxiv.org/abs/2210.06117}{{\ttfamily
  arXiv:2210.06117}} [astro-ph.CO]].

\bibitem{Bucko:2023eix}
J.~Bucko, S.~K. Giri, F.~H. Peters, and A.~Schneider, ``{\it {Probing the
  two-body decaying dark matter scenario with weak lensing and the cosmic
  microwave background}},''
  \href{http://dx.doi.org/10.1051/0004-6361/202347844}{{\em Astron. Astrophys.}
  {\bfseries 683} (2024) A152},
  [\href{http://arxiv.org/abs/2307.03222}{{\ttfamily arXiv:2307.03222}}
  [astro-ph.CO]].

\bibitem{Fuss:2024dam}
L.~Fu\ss{}, M.~Garny, and A.~Ibarra, ``{\it {Minimal decaying dark matter: from
  cosmological tensions to neutrino signatures}},''
  [\href{http://arxiv.org/abs/2403.15543}{{\ttfamily arXiv:2403.15543}}
  [hep-ph]].

\bibitem{Froggatt:1978nt}
C.~D. Froggatt and H.~B. Nielsen, ``{\it {Hierarchy of Quark Masses, Cabibbo
  Angles and CP Violation}},''
  \href{http://dx.doi.org/10.1016/0550-3213(79)90316-X}{{\em Nucl. Phys. B}
  {\bfseries 147} (1979) 277--298}.

\bibitem{Leurer:1993gy}
M.~Leurer, Y.~Nir, and N.~Seiberg, ``{\it {Mass matrix models: The Sequel}},''
  \href{http://dx.doi.org/10.1016/0550-3213(94)90074-4}{{\em Nucl. Phys. B}
  {\bfseries 420} (1994) 468--504},
  [\href{http://arxiv.org/abs/hep-ph/9310320}{{\ttfamily
  arXiv:hep-ph/9310320}}].

\bibitem{Leurer:1992wg}
M.~Leurer, Y.~Nir, and N.~Seiberg, ``{\it {Mass matrix models}},''
  \href{http://dx.doi.org/10.1016/0550-3213(93)90112-3}{{\em Nucl. Phys. B}
  {\bfseries 398} (1993) 319--342},
  [\href{http://arxiv.org/abs/hep-ph/9212278}{{\ttfamily
  arXiv:hep-ph/9212278}}].

\bibitem{Calibbi:2015sfa}
L.~Calibbi, A.~Crivellin, and B.~Zald\'\i{}var, ``{\it {Flavor portal to dark
  matter}},'' \href{http://dx.doi.org/10.1103/PhysRevD.92.016004}{{\em Phys.
  Rev. D} {\bfseries 92} no.~1, (2015) 016004},
  [\href{http://arxiv.org/abs/1501.07268}{{\ttfamily arXiv:1501.07268}}
  [hep-ph]].

\bibitem{Cheek:2022yof}
A.~Cheek, J.~K. Osi\'nski, L.~Roszkowski, and S.~Trojanowski, ``{\it {Dark
  matter production through a non-thermal flavon portal}},''
  \href{http://dx.doi.org/10.1007/JHEP03(2023)149}{{\em JHEP} {\bfseries 03}
  (2023) 149}, [\href{http://arxiv.org/abs/2211.02057}{{\ttfamily
  arXiv:2211.02057}} [hep-ph]].

\bibitem{Babu:2023zni}
K.~S. Babu, S.~Chakdar, N.~Das, D.~K. Ghosh, and P.~Ghosh, ``{\it {FIMP dark
  matter from flavon portals}},''
  \href{http://dx.doi.org/10.1007/JHEP07(2023)143}{{\em JHEP} {\bfseries 07}
  (2023) 143}, [\href{http://arxiv.org/abs/2305.03167}{{\ttfamily
  arXiv:2305.03167}} [hep-ph]].

\bibitem{Mandal:2023jnv}
R.~Mandal and T.~Tong, ``{\it {Exploring freeze-out and freeze-in dark matter
  via effective Froggatt-Nielsen theory}},''
  \href{http://dx.doi.org/10.1088/1475-7516/2023/11/074}{{\em JCAP} {\bfseries
  11} (2023) 074}, [\href{http://arxiv.org/abs/2307.14972}{{\ttfamily
  arXiv:2307.14972}} [hep-ph]].

\bibitem{CMB-S4:2016ple}
{\bfseries CMB-S4} Collaboration, K.~N. Abazajian {\em et~al.}, ``{\it {CMB-S4
  Science Book, First Edition}},''
  [\href{http://arxiv.org/abs/1610.02743}{{\ttfamily arXiv:1610.02743}}
  [astro-ph.CO]].

\bibitem{CMB-HD:2022bsz}
{\bfseries CMB-HD} Collaboration, S.~Aiola {\em et~al.}, ``{\it {Snowmass2021
  CMB-HD White Paper}},'' [\href{http://arxiv.org/abs/2203.05728}{{\ttfamily
  arXiv:2203.05728}} [astro-ph.CO]].

\bibitem{Banks:2010zn}
T.~Banks and N.~Seiberg, ``{\it {Symmetries and Strings in Field Theory and
  Gravity}},'' \href{http://dx.doi.org/10.1103/PhysRevD.83.084019}{{\em Phys.
  Rev. D} {\bfseries 83} (2011) 084019},
  [\href{http://arxiv.org/abs/1011.5120}{{\ttfamily arXiv:1011.5120}}
  [hep-th]].

\bibitem{Reece:2023czb}
M.~Reece, ``{\it {TASI Lectures: (No) Global Symmetries to Axion Physics}},''
  \href{http://dx.doi.org/10.22323/1.439.0008}{{\em PoS} {\bfseries TASI2022}
  (2024) 008}, [\href{http://arxiv.org/abs/2304.08512}{{\ttfamily
  arXiv:2304.08512}} [hep-ph]].

\bibitem{Calibbi:2016hwq}
L.~Calibbi, F.~Goertz, D.~Redigolo, R.~Ziegler, and J.~Zupan, ``{\it {Minimal
  axion model from flavor}},''
  \href{http://dx.doi.org/10.1103/PhysRevD.95.095009}{{\em Phys. Rev. D}
  {\bfseries 95} no.~9, (2017) 095009},
  [\href{http://arxiv.org/abs/1612.08040}{{\ttfamily arXiv:1612.08040}}
  [hep-ph]].

\bibitem{Ema:2016ops}
Y.~Ema, K.~Hamaguchi, T.~Moroi, and K.~Nakayama, ``{\it {Flaxion: a minimal
  extension to solve puzzles in the standard model}},''
  \href{http://dx.doi.org/10.1007/JHEP01(2017)096}{{\em JHEP} {\bfseries 01}
  (2017) 096}, [\href{http://arxiv.org/abs/1612.05492}{{\ttfamily
  arXiv:1612.05492}} [hep-ph]].

\bibitem{Qiu:2023igq}
Y.-C. Qiu, J.-W. Wang, and T.~T. Yanagida, ``{\it {Predictions of mee and
  neutrino mass from a consistent Froggatt-Nielsen model}},''
  \href{http://dx.doi.org/10.1103/PhysRevD.108.115021}{{\em Phys. Rev. D}
  {\bfseries 108} no.~11, (2023) 115021},
  [\href{http://arxiv.org/abs/2307.16470}{{\ttfamily arXiv:2307.16470}}
  [hep-ph]].

\bibitem{Fedele:2020fvh}
M.~Fedele, A.~Mastroddi, and M.~Valli, ``{\it {Minimal Froggatt-Nielsen
  textures}},'' \href{http://dx.doi.org/10.1007/JHEP03(2021)135}{{\em JHEP}
  {\bfseries 03} (2021) 135},
  [\href{http://arxiv.org/abs/2009.05587}{{\ttfamily arXiv:2009.05587}}
  [hep-ph]].

\bibitem{Hui:2021tkt}
L.~Hui, ``{\it {Wave Dark Matter}},''
  \href{http://dx.doi.org/10.1146/annurev-astro-120920-010024}{{\em Ann. Rev.
  Astron. Astrophys.} {\bfseries 59} (2021) 247--289},
  [\href{http://arxiv.org/abs/2101.11735}{{\ttfamily arXiv:2101.11735}}
  [astro-ph.CO]].

\bibitem{Lella:2023bfb}
A.~Lella, P.~Carenza, G.~Co', G.~Lucente, M.~Giannotti, A.~Mirizzi, and
  T.~Rauscher, ``{\it {Getting the most on supernova axions}},''
  \href{http://dx.doi.org/10.1103/PhysRevD.109.023001}{{\em Phys. Rev. D}
  {\bfseries 109} no.~2, (2024) 023001},
  [\href{http://arxiv.org/abs/2306.01048}{{\ttfamily arXiv:2306.01048}}
  [hep-ph]].

\bibitem{Adams:2022pbo}
C.~B. Adams {\em et~al.}, ``{\it {Axion Dark Matter}},'' in {\em {Snowmass
  2021}}.
\newblock 3, 2022.
\newblock [\href{http://arxiv.org/abs/2203.14923}{{\ttfamily arXiv:2203.14923}}
  [hep-ex]].

\bibitem{Calore:2022pks}
F.~Calore, A.~Dekker, P.~D. Serpico, and T.~Siegert, ``{\it {Constraints on
  light decaying dark matter candidates from 16~yr of INTEGRAL/SPI
  observations}},'' \href{http://dx.doi.org/10.1093/mnras/stad457}{{\em Mon.
  Not. Roy. Astron. Soc.} {\bfseries 520} no.~3, (2023) 4167--4172},
  [\href{http://arxiv.org/abs/2209.06299}{{\ttfamily arXiv:2209.06299}}
  [hep-ph]].

\bibitem{Gondolo:1990dk}
P.~Gondolo and G.~Gelmini, ``{\it {Cosmic abundances of stable particles:
  Improved analysis}},''
  \href{http://dx.doi.org/10.1016/0550-3213(91)90438-4}{{\em Nucl. Phys. B}
  {\bfseries 360} (1991) 145--179}.

\bibitem{Drees:2015exa}
M.~Drees, F.~Hajkarim, and E.~R. Schmitz, ``{\it {The Effects of QCD Equation
  of State on the Relic Density of WIMP Dark Matter}},''
  \href{http://dx.doi.org/10.1088/1475-7516/2015/06/025}{{\em JCAP} {\bfseries
  06} (2015) 025}, [\href{http://arxiv.org/abs/1503.03513}{{\ttfamily
  arXiv:1503.03513}} [hep-ph]].

\bibitem{Lillard:2018zts}
B.~Lillard, M.~Ratz, T.~Tait, M.~P., and S.~Trojanowski, ``{\it {The Flavor of
  Cosmology}},'' \href{http://dx.doi.org/10.1088/1475-7516/2018/07/056}{{\em
  JCAP} {\bfseries 07} (2018) 056},
  [\href{http://arxiv.org/abs/1804.03662}{{\ttfamily arXiv:1804.03662}}
  [hep-ph]].

\bibitem{McDonald:2001vt}
J.~McDonald, ``{\it {Thermally generated gauge singlet scalars as
  selfinteracting dark matter}},''
  \href{http://dx.doi.org/10.1103/PhysRevLett.88.091304}{{\em Phys. Rev. Lett.}
  {\bfseries 88} (2002) 091304},
  [\href{http://arxiv.org/abs/hep-ph/0106249}{{\ttfamily
  arXiv:hep-ph/0106249}}].

\bibitem{Kusenko:2006rh}
A.~Kusenko, ``{\it {Sterile neutrinos, dark matter, and the pulsar velocities
  in models with a Higgs singlet}},''
  \href{http://dx.doi.org/10.1103/PhysRevLett.97.241301}{{\em Phys. Rev. Lett.}
  {\bfseries 97} (2006) 241301},
  [\href{http://arxiv.org/abs/hep-ph/0609081}{{\ttfamily
  arXiv:hep-ph/0609081}}].

\bibitem{Petraki:2007gq}
K.~Petraki and A.~Kusenko, ``{\it {Dark-matter sterile neutrinos in models with
  a gauge singlet in the Higgs sector}},''
  \href{http://dx.doi.org/10.1103/PhysRevD.77.065014}{{\em Phys. Rev. D}
  {\bfseries 77} (2008) 065014},
  [\href{http://arxiv.org/abs/0711.4646}{{\ttfamily arXiv:0711.4646}}
  [hep-ph]].

\bibitem{Hall:2009bx}
L.~J. Hall, K.~Jedamzik, J.~March-Russell, and S.~M. West, ``{\it {Freeze-In
  Production of FIMP Dark Matter}},''
  \href{http://dx.doi.org/10.1007/JHEP03(2010)080}{{\em JHEP} {\bfseries 03}
  (2010) 080}, [\href{http://arxiv.org/abs/0911.1120}{{\ttfamily
  arXiv:0911.1120}} [hep-ph]].

\bibitem{Bernal:2017kxu}
N.~Bernal, M.~Heikinheimo, T.~Tenkanen, K.~Tuominen, and V.~Vaskonen, ``{\it
  {The Dawn of FIMP Dark Matter: A Review of Models and Constraints}},''
  \href{http://dx.doi.org/10.1142/S0217751X1730023X}{{\em Int. J. Mod. Phys. A}
  {\bfseries 32} no.~27, (2017) 1730023},
  [\href{http://arxiv.org/abs/1706.07442}{{\ttfamily arXiv:1706.07442}}
  [hep-ph]].

\bibitem{Bauer:2017qwy}
M.~Bauer and T.~Plehn, \href{http://dx.doi.org/10.1007/978-3-030-16234-4}{{\em
  {Yet Another Introduction to Dark Matter}: {The Particle Physics Approach}}},
  vol.~959 of {\em Lecture Notes in Physics}.
\newblock Springer, 2019.
\newblock [\href{http://arxiv.org/abs/1705.01987}{{\ttfamily arXiv:1705.01987}}
  [hep-ph]].

\bibitem{ParticleDataGroup:2022pth}
{\bfseries Particle Data Group} Collaboration, R.~L. Workman {\em et~al.},
  ``{\it {Review of Particle Physics}},''
  \href{http://dx.doi.org/10.1093/ptep/ptac097}{{\em PTEP} {\bfseries 2022}
  (2022) 083C01}.

\bibitem{Turner:1986tb}
M.~S. Turner, ``{\it {Thermal Production of Not SO Invisible Axions in the
  Early Universe}},'' \href{http://dx.doi.org/10.1103/PhysRevLett.59.2489}{{\em
  Phys. Rev. Lett.} {\bfseries 59} (1987) 2489}. [Erratum: Phys.Rev.Lett. 60,
  1101 (1988)].

\bibitem{Salvio:2013iaa}
A.~Salvio, A.~Strumia, and W.~Xue, ``{\it {Thermal axion production}},''
  \href{http://dx.doi.org/10.1088/1475-7516/2014/01/011}{{\em JCAP} {\bfseries
  01} (2014) 011}, [\href{http://arxiv.org/abs/1310.6982}{{\ttfamily
  arXiv:1310.6982}} [hep-ph]].

\bibitem{DEramo:2021lgb}
F.~D'Eramo, F.~Hajkarim, and S.~Yun, ``{\it {Thermal QCD Axions across
  Thresholds}},'' \href{http://dx.doi.org/10.1007/JHEP10(2021)224}{{\em JHEP}
  {\bfseries 10} (2021) 224},
  [\href{http://arxiv.org/abs/2108.05371}{{\ttfamily arXiv:2108.05371}}
  [hep-ph]].

\bibitem{Bouzoud:2024bom}
K.~Bouzoud and J.~Ghiglieri, ``{\it {Thermal axion production at hard and soft
  momenta}},'' [\href{http://arxiv.org/abs/2404.06113}{{\ttfamily
  arXiv:2404.06113}} [hep-ph]].

\bibitem{Mangano:2001iu}
G.~Mangano, G.~Miele, S.~Pastor, and M.~Peloso, ``{\it {A Precision calculation
  of the effective number of cosmological neutrinos}},''
  \href{http://dx.doi.org/10.1016/S0370-2693(02)01622-2}{{\em Phys. Lett. B}
  {\bfseries 534} (2002) 8--16},
  [\href{http://arxiv.org/abs/astro-ph/0111408}{{\ttfamily
  arXiv:astro-ph/0111408}}].

\bibitem{deSalas:2016ztq}
P.~F. de~Salas and S.~Pastor, ``{\it {Relic neutrino decoupling with flavour
  oscillations revisited}},''
  \href{http://dx.doi.org/10.1088/1475-7516/2016/07/051}{{\em JCAP} {\bfseries
  07} (2016) 051}, [\href{http://arxiv.org/abs/1606.06986}{{\ttfamily
  arXiv:1606.06986}} [hep-ph]].

\bibitem{Gariazzo:2019gyi}
S.~Gariazzo, P.~F. de~Salas, and S.~Pastor, ``{\it {Thermalisation of sterile
  neutrinos in the early Universe in the 3+1 scheme with full mixing
  matrix}},'' \href{http://dx.doi.org/10.1088/1475-7516/2019/07/014}{{\em JCAP}
  {\bfseries 07} (2019) 014},
  [\href{http://arxiv.org/abs/1905.11290}{{\ttfamily arXiv:1905.11290}}
  [astro-ph.CO]].

\bibitem{Akita:2020szl}
K.~Akita and M.~Yamaguchi, ``{\it {A precision calculation of relic neutrino
  decoupling}},'' \href{http://dx.doi.org/10.1088/1475-7516/2020/08/012}{{\em
  JCAP} {\bfseries 08} (2020) 012},
  [\href{http://arxiv.org/abs/2005.07047}{{\ttfamily arXiv:2005.07047}}
  [hep-ph]].

\bibitem{Froustey:2020mcq}
J.~Froustey, C.~Pitrou, and M.~C. Volpe, ``{\it {Neutrino decoupling including
  flavour oscillations and primordial nucleosynthesis}},''
  \href{http://dx.doi.org/10.1088/1475-7516/2020/12/015}{{\em JCAP} {\bfseries
  12} (2020) 015}, [\href{http://arxiv.org/abs/2008.01074}{{\ttfamily
  arXiv:2008.01074}} [hep-ph]].

\bibitem{Bennett:2020zkv}
J.~J. Bennett, G.~Buldgen, P.~F. De~Salas, M.~Drewes, S.~Gariazzo, S.~Pastor,
  and Y.~Y.~Y. Wong, ``{\it {Towards a precision calculation of $N_{\rm eff}$
  in the Standard Model II: Neutrino decoupling in the presence of flavour
  oscillations and finite-temperature QED}},''
  \href{http://dx.doi.org/10.1088/1475-7516/2021/04/073}{{\em JCAP} {\bfseries
  04} (2021) 073}, [\href{http://arxiv.org/abs/2012.02726}{{\ttfamily
  arXiv:2012.02726}} [hep-ph]].

\bibitem{Drewes:2024wbw}
M.~Drewes, Y.~Georis, M.~Klasen, L.~P. Wiggering, and Y.~Y.~Y. Wong, ``{\it
  {Towards a precision calculation of N $_{eff}$ in the Standard Model. Part
  III. Improved estimate of NLO contributions to the collision integral}},''
  \href{http://dx.doi.org/10.1088/1475-7516/2024/06/032}{{\em JCAP} {\bfseries
  06} (2024) 032}, [\href{http://arxiv.org/abs/2402.18481}{{\ttfamily
  arXiv:2402.18481}} [hep-ph]].

\bibitem{Rubakov:2017xzr}
V.~A. Rubakov and D.~S. Gorbunov, \href{http://dx.doi.org/10.1142/10447}{{\em
  {Introduction to the Theory of the Early Universe}: {Hot big bang theory}}}.
\newblock World Scientific, Singapore, 2017.

\bibitem{Viel:2005qj}
M.~Viel, J.~Lesgourgues, M.~G. Haehnelt, S.~Matarrese, and A.~Riotto, ``{\it
  {Constraining warm dark matter candidates including sterile neutrinos and
  light gravitinos with WMAP and the Lyman-alpha forest}},''
  \href{http://dx.doi.org/10.1103/PhysRevD.71.063534}{{\em Phys. Rev. D}
  {\bfseries 71} (2005) 063534},
  [\href{http://arxiv.org/abs/astro-ph/0501562}{{\ttfamily
  arXiv:astro-ph/0501562}}].

\bibitem{Boyarsky:2008xj}
A.~Boyarsky, J.~Lesgourgues, O.~Ruchayskiy, and M.~Viel, ``{\it {Lyman-alpha
  constraints on warm and on warm-plus-cold dark matter models}},''
  \href{http://dx.doi.org/10.1088/1475-7516/2009/05/012}{{\em JCAP} {\bfseries
  05} (2009) 012}, [\href{http://arxiv.org/abs/0812.0010}{{\ttfamily
  arXiv:0812.0010}} [astro-ph]].

\bibitem{Murgia:2017lwo}
R.~Murgia, A.~Merle, M.~Viel, M.~Totzauer, and A.~Schneider, ``{\it
  {''Non-cold'' dark matter at small scales: a general approach}},''
  \href{http://dx.doi.org/10.1088/1475-7516/2017/11/046}{{\em JCAP} {\bfseries
  11} (2017) 046}, [\href{http://arxiv.org/abs/1704.07838}{{\ttfamily
  arXiv:1704.07838}} [astro-ph.CO]].

\bibitem{Drewes:2016upu}
M.~Drewes {\em et~al.}, ``{\it {A White Paper on keV Sterile Neutrino Dark
  Matter}},'' \href{http://dx.doi.org/10.1088/1475-7516/2017/01/025}{{\em JCAP}
  {\bfseries 01} (2017) 025},
  [\href{http://arxiv.org/abs/1602.04816}{{\ttfamily arXiv:1602.04816}}
  [hep-ph]].

\bibitem{Garny:2018ali}
M.~Garny and J.~Heisig, ``{\it {Interplay of super-WIMP and freeze-in
  production of dark matter}},''
  \href{http://dx.doi.org/10.1103/PhysRevD.98.095031}{{\em Phys. Rev. D}
  {\bfseries 98} no.~9, (2018) 095031},
  [\href{http://arxiv.org/abs/1809.10135}{{\ttfamily arXiv:1809.10135}}
  [hep-ph]].

\bibitem{Decant:2021mhj}
Q.~Decant, J.~Heisig, D.~C. Hooper, and L.~Lopez-Honorez, ``{\it
  {Lyman-\ensuremath{\alpha} constraints on freeze-in and superWIMPs}},''
  \href{http://dx.doi.org/10.1088/1475-7516/2022/03/041}{{\em JCAP} {\bfseries
  03} (2022) 041}, [\href{http://arxiv.org/abs/2111.09321}{{\ttfamily
  arXiv:2111.09321}} [astro-ph.CO]].

\bibitem{Ringwald:2004np}
A.~Ringwald and Y.~Y.~Y. Wong, ``{\it {Gravitational clustering of relic
  neutrinos and implications for their detection}},''
  \href{http://dx.doi.org/10.1088/1475-7516/2004/12/005}{{\em JCAP} {\bfseries
  12} (2004) 005}, [\href{http://arxiv.org/abs/hep-ph/0408241}{{\ttfamily
  arXiv:hep-ph/0408241}}].

\bibitem{Weinberg:1979sa}
S.~Weinberg, ``{\it {Baryon and Lepton Nonconserving Processes}},''
  \href{http://dx.doi.org/10.1103/PhysRevLett.43.1566}{{\em Phys. Rev. Lett.}
  {\bfseries 43} (1979) 1566--1570}.

\bibitem{Yanagida:1979as}
T.~Yanagida, ``{\it {Horizontal gauge symmetry and masses of neutrinos}},''
  {\em Conf. Proc. C} {\bfseries 7902131} (1979) 95--99.

\end{thebibliography}\endgroup

\end{document}